# Effect of fatigue on the intra-cycle acceleration in front crawl swimming: A time-frequency analysis

V. Tella[1], J. L. Toca-Herrera[2*], J. E. Gallach[1], J. Benavent[1], L. M. González[1], R. Arellano[3]

[1]Universidad de Valencia; Departamento de Educación Física

[2] Biosurfaces unit, BiomaGUNE

[3]Universidad de Granada; Departamento de Educación Física

**\*Corresponding author:**

address: Biosurfaces Unit, CIC BiomaGUNE, Paseo Miramón 182, 20009 San Sebastian, Spain

telephone: +34 943 00 53 13

fax numbers: +34 943 00 53 14

E-mail: jltocaherrera@cicbiomagune.es

# ABSTRACT

The present study analyzes the changes in acceleration produced by swimmers before and after fatiguing effort.

The subjects (n=15) performed a 25-meter crawl series at maximum speed without fatigue, and a second series with fatigue. The data were registered with a synchronized system that consisted in a position transducer (1 kHz) and a video photogrametry (50Hz). The acceleration ($ms^{-2}$) was obtained by the derivative analysis of the variation of the position with time. The amplitude in the time domain was calculated with the root mean square (RMS); while the peak power (PP), the peak power frequency (PPF) and the spectrum area (SA) was calculated in the frequency domain with Fourier analysis.

On one hand, the results of the temporal domain show that the RMS change percentage between series was 67.5% ($p<0.001$). On the other hand, PP, PPF, and SA show significant changes ($p<0.001$). PP and SA were reduced by 63.1% and 59.5%, respectively. Our results show that the acceleration analysis of the swimmer with Fourier analysis permits a more precise understanding of which propulsive forces contribute to the swimmer performance before and after fatigue appears.

## INTRODUCTION

Swimming displacement is the result of the application of propulsive force to overcome the body drag (Maglischo et al., 1987; Costill et al., 1987; Chengalur and Brown, 1992). In this sense, the acceleration that a swimmer achieves while moving will be the result of the interaction between the aforementioned forces and the water. Indeed, a right body position and adequate both arm and leg movements will influence a more efficient front crawl technique (Schleihauf et al., 1986; Maglischo et al., 1988; Toussaint et al., 1988; Chatard et al., 1990). Some recent theoretical models (Bixler and Riewald, 2002; Rouboa et al., 2006) show the importance of making accelerated propulsive movements to increase drag forces. Conversely, if acceleration lessens while the propulsive movements are being made, the magnitude of the produced drag force diminishes (Bixler, 2005).

Specifically, front crawl swimmers carry out three underwater phases with their arms *(Downsweep, insweep and upsweep)* as well as a series of leg movements to generate propulsion. The most frequent leg-arm coordination is that which includes six kicks per cycle (Maglischo, 1993). As in the aforementioned theoretical models, in 1983 Schleihauf and co-workers had already shown that the various arm propulsive phases made by top swimmers were done so in an accelerated fashion.

In contrast, specific movements that are characteristic of the front crawl style, such as body roll, how the hands enter the water, breathing and the depth that certain parts of the body reach, may bring about decelerations if they are performed incorrectly. On the other hand, when rolling is synchronised with the propulsive movements that the arms and legs perform, such movements are more efficient and produce greater accelerations (Toussaint and Beek, 1992; Yanai, 2001).

Traditionally stroke rate and length have been used as parameters to determine swimming speed. These parameters vary as a consequence of the distance of the event and the level of fatigue accumulated during this event. In relation to fatigue, reduced propulsive forces and an increased resistance to forward movement results in loss of speed (Di Prampero et al., 1974). Specifically, this reduction is produced by shortening the stroke length and by the difficulty to produce higher stroke rates (Keskinen, 1997).

The coordination of propulsive arm movements seems to be affected by fatigue. In this sense, Alberty and co-workers (2003), who used the Index of Coordination as described by Chollet (2000), observed a significant reduction of this index.

Intra-cycle velocity has been analyzed in several studies calculating the position related to time. However, the intra-cycle acceleration has not been studied as variable in the reviewed literature. The main purpose of this study is to analyze the acceleration produced during the front crawl swimming stroke cycle, differentiating time and frequency domains. A practical application is included, observing how fatigue affects intra-cycle acceleration.

## MATERIALS AND METHODS

*Subjects*

After having signed an informed consent, sixteen regional and national swimmers (mean ± standard error of the mean, SEM; age 17.0 ± 0.8 years; weight, 69.4 ± 2.5; height 173.0 ±1.6 cm) took part in the research study. The swimmers neither suffered musculoskeletal pathologies nor restrictions, which hindered their performance during events.

*Experimental procedure*

The study was conducted in two stages. In the first stage all the subjects performed several trials to get familiar with the equipment and the experimental organization. A reproducibility test was performed for this protocol by taking two individual measurements with a 48-hours interval between them. Intra-class correlation coefficients of the tests revealed good test re-test reliability (range 0.83-0.89). Some days later two trials covering 25m front crawl sprint with water start were performed, including at least 10 minutes rest between trials. The best 25m-time was selected for the intra-cycle analysis. In the second stage, two days later, the subjects performed a 75m front crawl repetition at maximum intensity; after this effort and during the next 10 s the recording system was attached to the swimmer's belt to perform a 25m sprint under fatigue conditions.

*Measuring acceleration*

Acceleration was obtained from the position-time data recorded using a position transducer (ISOcontrol, ATE Micro, Madrid, Spain), recording at 1kHz. The apparatus consisted in a resistive sensor (i.e., which produced a resistance of 250 g) with a coiled cable that was fastened to the swimmers' waists at the height of second and third lumbar vertebrae by means of a belt. The subjects did the test sets inside the swimming pool.

All the pre-test and post-test data were registered and converted from analogical to digital (12-bit; DAQCard–700; National Instrument, Austin, USA). Data were stored on a hard disk for subsequent analyses.

*Underwater videography*

A full stroke cycle was recorded using an underwater video camera, perpendicular to the swimmer's plane of displacement (the signal was registered at 50 Hz).

*Data analysis*

To analyse the position signals, a specific program was written and run in Matlab 7.1 (R14) (Mathworks Inc., Natick, USA). The position signal was derived two times to obtain the corresponding acceleration signal ($ms^{-2}$). The acceleration signal was filtered to preserve only those frequencies of interest for the study. A Butterworth fourth-order digital filter was used for this purpose with a band-pass of 1-20 Hz. This signal was then analysed in both the time and frequency domains. Given the fact that the size of the acceleration was unstable in the first and final seconds of each swimming set, the 5 central seconds were selected in each trial to analyse the signal (Caty et al., 2006).

The signal amplitude was examined in the time domain with a root mean square (RMS), and processed in 100 ms-sized blocks. Likewise, the coefficient of variation (CV) of the acceleration signal was calculated in accordance with the following equation: CV = (100 x $\sigma$ x $\mu^{-1}$); were $\sigma$ is the ratio of the standard deviation, $\mu$ is the mean, and the result are expressed in a percentage (%).

The frequency spectrum amplitude was analysed with the periodogram method (Pollock, 1999), which permits to discover the hidden frequencies in a signal[1]. This was performed by using the Matlab SPECTRUM function, and averaged with the Welch method. A 1024-point Hamming window was used for this purpose. The dependent variables calculated in the frequency domain were: the peak power (PP, the highest value of the power spectrum), the peak power frequency (PPF, the frequency associated with the peak power) and the total power contained in the spectrum area (SA), which is the total power of the whole spectrum between 0 and 20 Hz. Figure 1 shows an example of the analysis carried out.

The images captured on video were used to obtain data for the swimming sets. Photograms corresponding to the start and end of each set were visually marked. These two events were used to calculate the space-time data corresponding to a swimming series. The mean velocity, the stroke frequency and the stroke length were obtained with the data of the position in time of the selected swimming set. In addition, the index of coordination was calculated following the protocols according to Chollet (2000).

*Statistical analysis*

The statistical analysis was performed with the SPSS software, version 12.0 (SPSS Inc., Chicago, IL, USA). The assumed normality (K-S normality test) was verified for all the variables prior to doing all the analyses.

To obtain the descriptive statistics (mean, standard mean error), standard statistical methods were used. A Student's t-test for the related samples to establish differences between the values obtained before and after the fatigue protocol was applied.

In order to establish relationships between variables, a Pearson's correlation coefficient (*r*) was used in both the pre-test and post-test (*n*=15). Additionally, both trials were used to search for correlations (*n*=30). For all the analyses, those differences whose probability was below 5% ($p<0.05$), and due to random, were accepted.

---

[1] The periodogram considers all the frequencies and correlates each frequency with the data of the series in order to estimate the importance of a particular frequency in the series. It assigns to each frequency a value called *intensity* of frequency denoted by $I(w) = [a(w)]^2 + [b(w)]^2$, where $a(w) = \frac{2}{N} \sum_t y_t \cos(tw)$ and $b(w) = \frac{2}{N} \sum_t y_t \sin(tw)$. The expressions $a(w)$ and $b(w)$ can be seen as the covariance of the series with the cosine and sine functions (N: window length, t: time, w: frequency). Note that when N goes to infinity the expected value of the periodogram equals the true power spectral density of the signal.

# RESULTS

*Basic description of a stroke cycle*

The analysis results of a stroke cycle are provided in Table 1. The values of the mean speed and stroke frequency are significantly reduced ($p<0.001$), whereas no significant changes were seen for stroke length and the index of coordination after the fatigue protocol.

*Time domain analysis of the acceleration signal*

Table 2 displays the descriptive statistics of the acceleration signal produced during the pre-test and post-test. The percentage of change of the RMS was 67.51% ($p<0.001$). Significant changes were also noted for both the maximum and minimum values ($p<0.001$) after performing the fatigue protocol. The acceleration signal fluctuated similarly in relation to the average both before and after the fatigue protocol. The variation coefficient did not change significantly ($p=0.47$), and the results obtained were 78.28% and 76.76% in the pre-test and post-test, respectively.

*Frequency domain analysis of the acceleration signal*

Figure 2 illustrates the different frequency spectra obtained during the swimming trials without fatigue. For this, the spectrum of accelerations generated by each swimmer has been represented. Three types stand out and are indicated as models *a, b* and *c*.

The influence that fatigue has on the frequency domain variables is shown in Table 3, where significant changes are seen ($p<0,001$) in the three variables calculated (peak power, peak power frequency and spectra area). Specifically, the peak power and the spectra area were reduced by 63.11% and 59.53%, respectively.

*Relationship between variables*

The relationship between speed as a performance parameter and the different acceleration variables analysed in both the time and frequency ($n=15$) domains are displayed in Table 4. A positive correlation was also achieved between the peak power frequency (PPF) and the frequency cycle (FC) in the two sets tested: $r$=0.68 (p<0.01) without fatigue and $r$=0.70 post-test with fatigue (p<0.01). Thus one can observe a significant correlation when both swimming trials were attempted ($n$=30).

**DISCUSSION**

*Acceleration signal analysis*

The type of signal used in our work is similar to that described by other authors, which comes close to the three typical acceleration phases described by Ohgi and co-workers in 2002. This similarity is most likely due to the relevance of the propulsive arm actions in the total front crawl swimmer's movement.

Traditionally, the acceleration signal has been modelled by using statistical parameters such as average, range and the coefficient of variation (CV). In our study we used the RMS (also known as an efficient value) as the main statistic parameter to characterise the acceleration amplitude. Our data show a clear relationship between the values obtained in the RMS and the speed achieved by the subjects, thus, revealing the robustness of RMS in the characterisation of the swimmers' acceleration signals. All these variations in acceleration produced during the crawl stroke and their relationship with the speed have already been described in a previous work (Counsilman and Wasilack, 1982). In this work, the authors reported a high correlation between the acceleration of hands-arms and the swimmers' final speed. Nonetheless, some researchers (Toussaint et al., 2002; Rouboa et al., 2006) subsequently pointed out that this relationship masks a more complex phenomenon, accounting for the fact that the hands do not produce a constant acceleration from the beginning of the movement to the end, rather the swimmers' hands accelerate and decelerate in pulses. Although our results are not totally comparable with those works, some similarities exist. The fact that acceleration and deceleration are part of a normal swimming set is corroborated by the two calculated dispersion statistics (the maximun and the minimum appear to correlate with the velocity of our results).

On the other hand, the CV also appears to be cited in the scientific literature as a reliable statistic to distinguish the economy of swimming (Miyashita, 1970; Nigg, 1983; Alves et al., 1996). The coefficients in this work show a CV close to ~77% in the acceleration signal, and it is the only parameter that has shown no correlation with velocity. In this sense, the CV does not contribute with any type of information to our study that may discriminate a more or less economic swimming technique.

The second part of the analysis has focused on the frequency domain. To discriminate the signal into its different frequencies, a mathematical process was used based on the Fast Fourier Transform (FFT). The visual analysis of a graph, which corresponds to a non-periodical signal, does not normally provide information of the repeated cycles and their associated frequencies. Nonetheless, since swimming is a cyclic sport, the regularities that make up a swimming cycle have normally been analysed visually. In fact, the observation of graphs and the counting of peaks have even been used to characterise the various components of swimming (Maglischo, 1993). However, swimming signals are not (in general) periodic signals and they are treated mathematically through a Fast Fourier transform (FFT). In this way, the parameters that escape the visual perception of the researcher can be accurately verified.

In all the models, which appear in our work (Figure 2), a maximum peak power may be seen at which most of the energy produced by the swimmer accumulates. The frequency at which this peak occurs ranges from 4.39 Hz to 8.91 Hz. Depending on the model, the appearance of two or more peaks with lower values may be observed at different frequencies. Consequently in model *c*, the different oscillators that could potentially have an effect on peak amplitude are produced at a similar frequency (i.e. the different oscillators are seen to be coherent as far as frequency is concerned). In model *b* however, the action of two oscillators with different amplitudes and varying frequencies may be intuitively sensed. Thus, model *a* shows greater entropy than the others.

The visual analysis of the spectra shows, above all, a peak of maximum amplitude, which coincides in almost all the analysed models. Apparently, there is an oscillator that shows a frequency of around 5 Hz, which has led to a high percentage of acceleration produced during the signal. It seems obvious that this maximum peak may be due to accelerations that are fundamentally produced by the arms.

Specifically, the analysis performed before the images were taken shows that the average time used in each stroke cycle is approximately ~1.1 s, without fatigue, and around ~1.3 s with fatigue. These data are similar to those reported by other authors (Craig et al., 1985; Arellano et al., 1994; Tella et al., 2002) who found a stroke frequency of ~0.9 Hz during swimming. Consequently, if we take into account previous studies which describe 6 acceleration phases during front crawl swimming as a basis (Counsilman and Wasilack, 1982), the result obtained is a frequency of approximately 6 accelerations per second in the arms, or in other words, a frequency of 6 Hz. Taking into account the mean value of the swimmers´ cycle time, the analyzed peak power frequency (PPF) takes a value of around 5.9 Hz without fatigue and 5.1 with fatigue. Although most of the propulsive forces generated during swimming (~85-90%) correspond to arm movements (Hollander et al., 1988; Deschodt et al., 1999), we need to explain foot action as the second most important oscillator in our analyses. This work has its limitations. For example, only body acceleration has been measured, and we were unable to deduce accurately the contribution to acceleration by the feet. Nevertheless, we believe that foot action may appear to coordinate with arm action. It would be interesting to do an in-depth study into this premise in a future research work by measuring the acceleration of both arms and legs in a synchronised fashion. In this sense, a complete on the coherence in the frequency and phase of both signals can be carried out.

The correlation analysis reveals a significant relationship between the frequency of the stroke cycle and the frequency at which the maximum amplitude peak of the spectrum takes place in our work. In this sense, it seems that the PPF variable could be statistically valid to quantify the effectiveness of the swimmers' stroke frequency.

*Effects of fatigue on acceleration*

The mean velocity values decreased due to the fatigue, which accumulated during the experiment, 84.5%, similarly to cycle frequency, which also decreased up to 85.7%. These decreases agree with those obtained by Weiss et al. (1988), although no significant changes were seen for the cycle values (i.e. stroke length and index of coordination), which also occurred in other studies in which the swimmers were submitted to a fatigue protocol (Chollet et al., 2000; Alberty el al., 2003). The highest dimension of fatigue provoked in our study could account for the fact that the found variations took only place in the stroke frequency when compared with the aforementioned studies.

During the acceleration analysis in the time and frequency domains, a loss of all the analysis parameters also appears, except for the coefficient of variation (CV). The reduction shown in the time domain values, specifically in the frequency domain values PP, PPF and SA, indicates the different role of the propulsive forces under fatigue conditions.

On the whole conclusion, the data obtained in this work suggest that analysing the swimmer's body acceleration through mathematical processes (e.g. FFT) permits to know the sequence of propulsive forces during swimming more accurately, and specifically to learn how they are modified after a fatigue protocol.

**ACKNOWLEDGMENTS**

This work has been supported by the emergent research group grant of the University of Valencia (UV-AE-20041029). JLTH is a Ramón y Cajal Research Fellow and therefore he thanks the RyC Program of the Spanish Government.

**FIGURES**

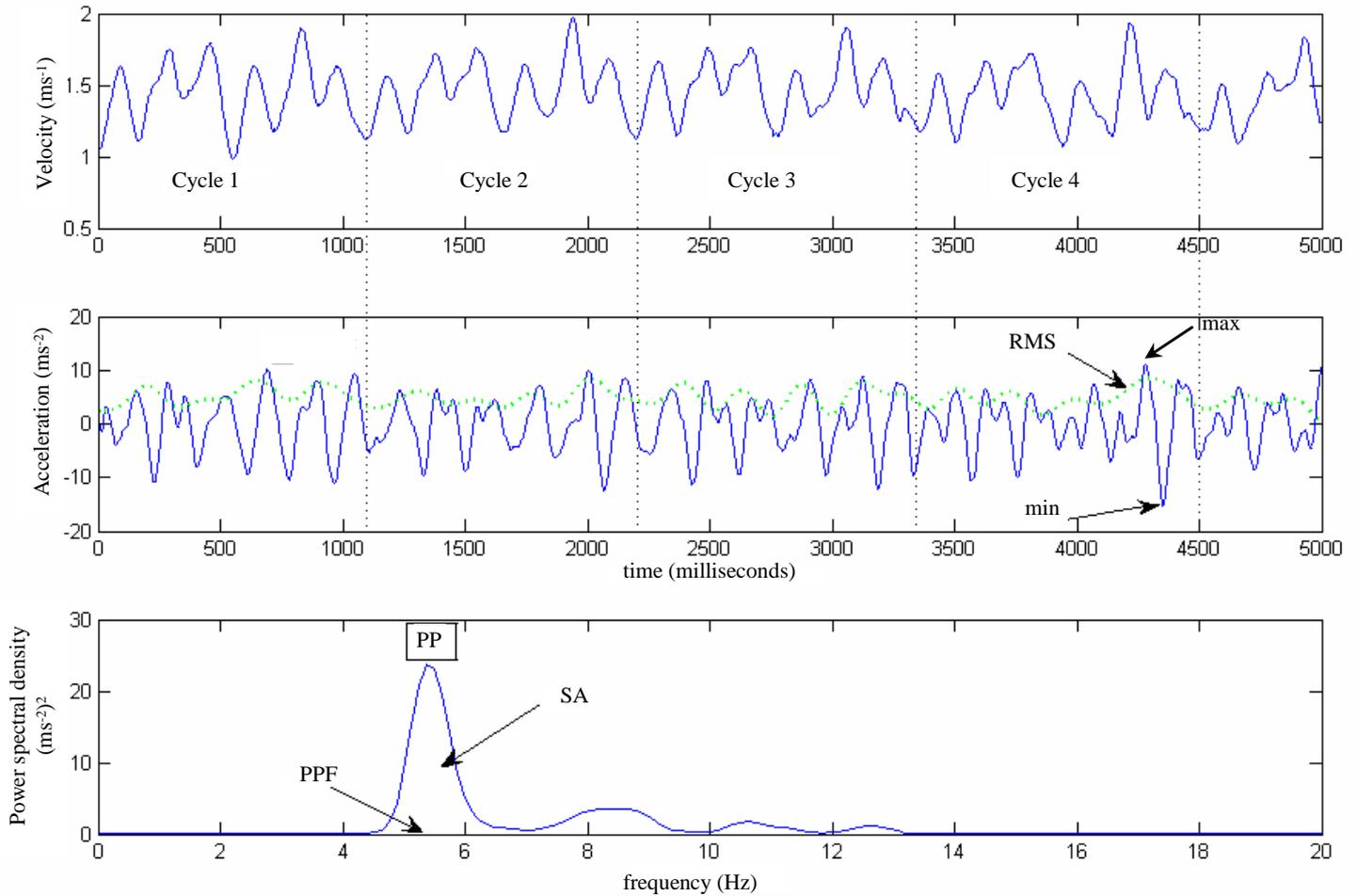

**Figure 1. Example of the analysis performed with the position signal in time.** In the upper panel we see the usual representation of an intra-cycle velocity. In the central panel we see our analysis proposal of the intra-cycle changes in accordance with acceleration. The original acceleration signal of a standard subject is depicted, as is the root mean square (RMS), and the maximum and minimum value at which acceleration takes place. In the lower panel we note the corresponding frequency analysis of the acceleration signal. It shows the peak power (PP), the peak power frequency (PPF) and the area below the spectrum curve (SA).

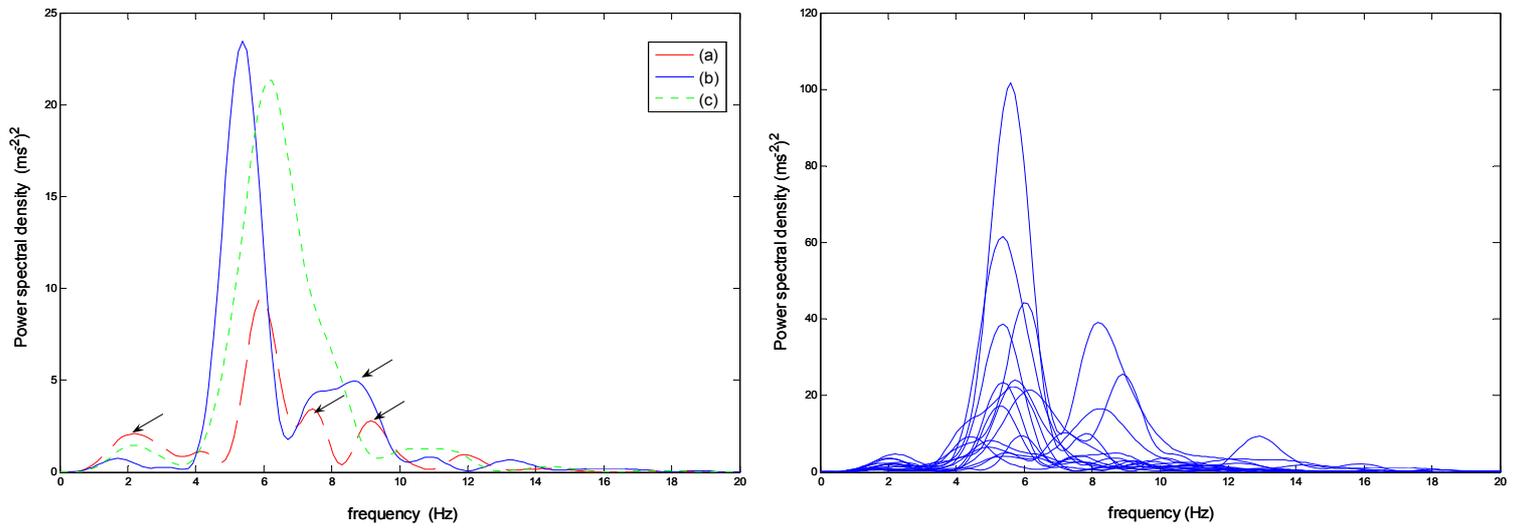

**Figure 2. Different spectra of the frequencies obtained during swimming trials without fatigue.** In the left-hand panel three standard spectra appear. The submaximum peaks are indicated with an arrow. Spectrum (a) shows different peak powers throughout the spectrum, and the frequency at which the maximum amplitude peak takes place is 5.98 Hz. Likewise three lower peaks are seen at: 2.11 Hz, 7.44 Hz and 9.13 Hz. Model (b) shows a peak power with a high amplitude and an associated frequency of 5.37 Hz, and there was only a second potentially lower peak at 8.6 Hz. Model (c) illustrates a spectrum with a single peak power at 6.22 Hz. In the right-hand panel we may see all the spectra obtained by the subjects.

**TABLES**

**Table 1. Descriptive statistics of a swimming cycle during the pre-test (without fatigue) and post-test (with fatigue).**

|  | Without fatigue (*n*=30) | With fatigue (*n*=30) |
|---|---|---|
| MV (ms$^{-1}$) | 1.60±0.04 | 1.34±0.4* |
| SF (cycles/min) | 55.26±1.20 | 47.36±1.71* |
| SL (meters/cycle) | 1.74±0.05 | 1.71±0.05 |
| IdC (%) | 5.49±0.67 | 7.06±1.10 |

*Data are expressed as mean±SEM. * indicates a significant difference respect without fatigue p<0.001. Mean velocity; MV, stroke frequency; SF, stroke length; SL and index of coordination, IdC.*

**Table 2. Analysis of acceleration in the time domain during the pre-test (without fatigue) and post-test (with fatigue).**

|  | Without fatigue (*n*=15) | With fatigue (*n*=15) |
|---|---|---|
| RMS (ms$^{-2}$) | 7.95±0.62 | 5.30±0.42* |
| Maximum (ms$^{-2}$) | 23.59±2.22 | 14.93±1.39* |
| Minimum (ms$^{-2}$) | −22.97±1.80 | −15.40±1.70* |

*Data are expressed as mean±SEM. * indicates a significant difference respect without fatigue p<0.001. RMS: root mean square.*

**Table 3. Frequency analysis of acceleration during the pre-test (without fatigue) and post-test (with fatigue).**

|  | Without fatigue (*n*=15) | With fatigue (*n*=15) |
|---|---|---|
| PP (ms$^{-2}$)$^2$ | 28.68±6.84 | 10.58±2.44* |
| PPF (Hz) | 5.88±0.31 | 5.12±0.36* |
| SA (ms$^{-2}$)$^2$ | 537.03±99.90 | 210.40±1.70* |

*Data are expressed as mean±SEM. * indicates a significant difference respect without fatigue p<0.001. PP; Peak Power, PPF; Peak Power Frequency, SA; Spectrum Area*

**Table 4. Correlation variables (r de Pearson), frequency analysis and swimming velocity**

| | Velocity without fatigue ($n$=15) | Velocity with fatigue ($n$=15) |
|---|---|---|
| RMS (ms$^{-2}$) | 0.68** | 0.75** |
| Maximum (ms$^{-2}$) | 0.53* | 0.59* |
| Minimum (ms$^{-2}$) | −0.54 * | −0.32 |
| PP (ms$^{-2}$)$^2$ | 0.45 | 0.65** |
| PPF (Hz) | 0.24 | 0.41 |
| SA (ms$^{-2}$)$^2$ | 0.53* | 0.75** |

*indicates $p<0.05$, ** indicates $p<0.01$. RMS; root mean square, PP; Peak Power, PPF; Peak Power Frequency, SA; Spectrum Area